\documentclass[12pt]{article}
\usepackage[dvips]{graphicx}
\begin{document}
\title{Massive charged scalar field in a Reissner-Nordstrom
black hole background: quasinormal ringing}
\author{R.A.Konoplya \\
Department of Physics, Dniepropetrovsk National University\\
St. Naukova 13, Dniepropetrovsk  49050, Ukraine\\
konoplya@ff.dsu.dp.ua}
\date{}
\maketitle
\thispagestyle{empty}
\begin{abstract}
We compute characteristic (quasinormal) frequencies corresponding
to decay of a massive charged scalar
field in a Reissner-Nordstrom black hole background.

It proves that,
contrary to the behavior at very late times,
at the stage of quasinormal ringing the
neutral perturbations  will damp slower than the charged ones.
In the limit of the extremal black hole the damping rate of charged and
neutral perturbations coincides. Possible connection of this with the
critical collapse in a massive scalar electrodynamics is discussed.

\end{abstract}

Black hole (BH) quasinormal modes (QNM's) govern the decay of
perturbations at intermediate times and are important when studying
dynamics of black holes  and external fields around them.
Now, the interest in quasinormal
modes is stipulated by three their features. The first is
connected with possibility of observing of QNM's, and of
obtaining the "footprint" of a black hole,  with the help
of a new generation of gravitational antennae which are under
construction (see \cite{Kokkotas-Schmidt} for a review).
The second concerns the Anti-de-Sitter/Conformal
Field Theory (AdS/CFT) correspondence \cite{Maldacena},
from which it follows that a
large black hole in AdS space corresponds to an approximately thermal
states in the CFT. Thus
the decay of the BH perturbations can be associated with the return
to thermal equilibrium of the perturbed state in CFT \cite{Horowitz1}.
Note that in AdS space, quasinormal modes of black holes govern the
decay of a field at late times as well, since there is no power-low
tails and the decay is always exponential (see \cite{Horowitz1} and
references therein).
The third feature is related to possible connection of BH QNM's in some
space-time geometries with the Choptuik scaling
\cite{Horowitz1}, \cite{Birmingham1}, \cite{Kim}, \cite{Konoplya3}.

When Horowitz and Hubeny calculated massless scalar quasinormal modes
for a
four-dimensional SAdS black hole they found that for intermediate
black holes, i.e. when the horizon radius is of order the
Anti-de Sitter radius (see also \cite{Konoplya1}), the
striking approximate relation takes place
\begin{equation}\label{01}
\omega_{Im} \sim \frac{1}{\gamma} r_{+},
\end{equation}
where $r_{+}$ is the horizon radius, $\gamma$ is the Choptuik
scaling parameter for a black hole being formed from a collapse of a
scalar field in free space. Then, exact connection between the
Choptuik scaling parameter and the imaginary part of the QNF's
was obtained for the
massive scalar field in the background of the three-dimensional BTZ
BH \cite{Birmingham1}. It gives us a hope that such a connection
between characteristic frequencies of black holes, at least in
special space-time geometries, and parameters describing situation
near the threshold formation of a black hole exists
for all black holes. Thus knowledge of such relations would be
a most useful, since calculation of quasinormal modes is much
simpler work then numerical simulation of gravitational collapse.

Then it was observed in \cite{Kim} that since the nearly
extremal RN BH background is effectively described by $AdS_{2}$ black
holes under a spherically symmetrical dimensional reduction
\cite{Strominger}, the relation between the scalar QN modes
of the nearly extremal RN BH and the Choptuik scaling may be obtained.

Recently it has been found that at the stage of quasinormal ringing the
neutral perturbations, corresponding to a  massless scalar field,
will damp slower than the charged ones for RN, RNAdS, and dialton black
holes \cite{Konoplya3}.
At very late times, on contrary, the charged perturbations will damp
slower, thereby dominating in a signal as was shown in \cite{Hod-Piran1}.
In addition the damping rates (imaginary part of $\omega$) of charged
and neutral quasinormal modes for massless scalar field coincide
for the nearly extreme RN black hole \cite{Konoplya3}.
This fact, although not understood, may be connected with
the Choptuik scaling if remembering that the Choptuik scaling parameter
$\gamma$ is the same ($0.37$) both for charged \cite{Hod-Piran} and neutral
\cite{Choptuik} massless scalar
field collapse.
The case of a massive scalar field is more complicated from a
critical collapse view. There both types of behavior appear
(see \cite{Gundlach} and references therein), I
(with a mass gap) and II (when the black hole being formed may be of
infinitesimal mass) depending on ratio of the length scale of the
initial data to the inverse Compton wavelength characteristic of the field.
Yet the critical collapse was studied only for
massless electrodynamics and for a massive scalar field.
The complete picture of critical collapse for massive scalar
electrodynamics is lacking as far as we are aware. This gives us one
of the reasons for studying the quasinormal spectrum associated with
decay of a charged massive scalar field.

The other reason is to find out how a massive scalar field,
interacting also electromagnetically with a charged black hole, decay
between an initial burst of radiation and the late time tails.
In this direction the evolution of a neutral massive scalar field
at late times was studied in \cite{Tomimatsu1} for a Schwarzshild
background, in \cite{Hod-Piran3}, \cite{Wang3}, \cite{Tomimatsu2}
for a RN background, and in \cite{Moderski-Rogatko} for the dilaton black
hole.
At the stage of quasinormal ringing,
decay of neutral massive scalar field was studied in \cite{Simone-Will}
for Schwarzshild and Kerr black holes.
The quasinormal modes of dilaton black holes were studied in
\cite{Konoplya3}, \cite{Piazza},
 \cite{Konoplya4}.

The complete picture of a charged massive scalar field evolution in
a charged black hole background is lacking at either stages.

We shall consider the evolution of the massive charged scalar
field in the background of the Reissner-Nordstrom metric:
\begin{equation}\label{1}
ds^{2}= -f(r) dt^{2} + f^{-1}(r)dr^{2} +r^{2}d\Omega^{2}_{2},
\end{equation}
where $ f(r)= 1-\frac{2 M}{r}+\frac{Q^2}{r^2}$. The wave equation of
the complex scalar field has the form:
\begin{equation}\label{2}
\phi_{;ab} g^{ab}-i e A_{a} g^{ab}(2 \phi_{; b}-i e A_{b}
\phi)-i e A_{a; b} g^{ab} \phi + \mu^2 \phi =0,
\end{equation}
here the electromagnetic potential $A_{t} = C -\frac{Q}{r}$, $C$ is a
constant.
After representation of the charged scalar field into spherical
harmonics the equation of motion takes the form:
\begin{equation}\label{3}
\psi_{,tt}+2\imath e \frac{Q}{r} \psi_{,t} - \psi_{, r^{*}r^{*}}+ V
\psi=0,
\end{equation}
where
\begin{equation}\label{4}
V = f(r)  \left(\frac{l(l+1)}{r^2} +\frac{2 M}{r^3} -
\frac{2 Q^2}{r^4} + \mu^2 \right) -e^{2} \frac{Q^2}{r^2},
\end{equation}
the tortoise coordinate is defined by the relation
$d r^*= \frac{d r}{f(r)}$, and $\psi = \psi (r) e^{-\imath \omega
t}$, $\omega = \omega_{Re} - i \omega_{Im}$.
The effective potential goes to a constant at the horizon and
at infinity.
We compute the quasinormal frequencies stipulated by the above
potential using the third order WKB formula of S.Iyer and C.Will
\cite{Will1}:
\begin{equation}\label{5}
\frac{\imath Q_{0}}{\sqrt{2 Q_{0}''}}
-\Lambda(n)-\Omega(n)=n+\frac{1}{2},
\end{equation}
where $\Lambda(n)$, $\Omega(n)$ are second and third order WKB
correction terms depending on the potential $Q$ and its derivatives
in the maximum. Here $Q = -V+\omega^{2} - 2 \frac{e Q}{r} \omega$.
Since $Q$ generally depends on $\omega$, the procedure of finding of
the QN frequencies is the following: one fixes all the parameters of
the QN frequency, namely, the multipole index $l$, the overtone
number $n$, of the black hole $Q$ and $M$, and of the field $m$ and
$e$; then one finds the value of $r=r_{0}$ at which $Q$ attains a maximum
as a numerical function of $\omega$ and substituting it into the
formula (\ref{5}) one finds $\omega$ which satisfies the equation
(\ref{5}). Note that the effective potential, being frequency
dependent, is, generally, complex. Nevertheless, following
\cite{Will1},
we treat $\omega$ as if it were real when finding $r_{0}$ and only
continue Eq.(\ref{5}) into the complex frequency plane. We deal here
with $n=0$ modes as those dominating in a signal.

First we examine the behaviour of the neutral massive scalar field in
a RN background. Recently in \cite{Wang3} through numerical simulation
of the wave equation solution it has been obtained that
the relaxation process depends on the value $M \mu$: when $M \mu \ll 1$
the relaxation depends on the field parameters and does not depend
on the spacetime parameters, while at $M \mu \gg 1 $ the dependence on the
black hole parameters appears. At the stage of quasinormal
oscillations the evolution of the massive
scalar field governed by QNF's  depends both on black hole and field
parameters.

\begin{figure}
\begin{center}
\includegraphics{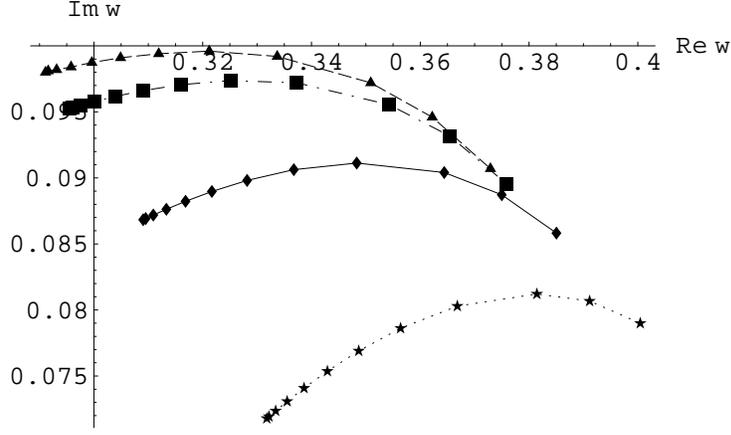}
\caption{RN QNF's for $l=1$, $n=0$, $\mu=0, 0.1, 0.2, 0.3$,
$Q$ runs  the values $0, .1, .2, .3, .4, .5, .6, .7, .8, .9, .95, .99$}
\label{plb1fig1}
\end{center}
\end{figure}

\begin{figure}
\begin{center}
\includegraphics{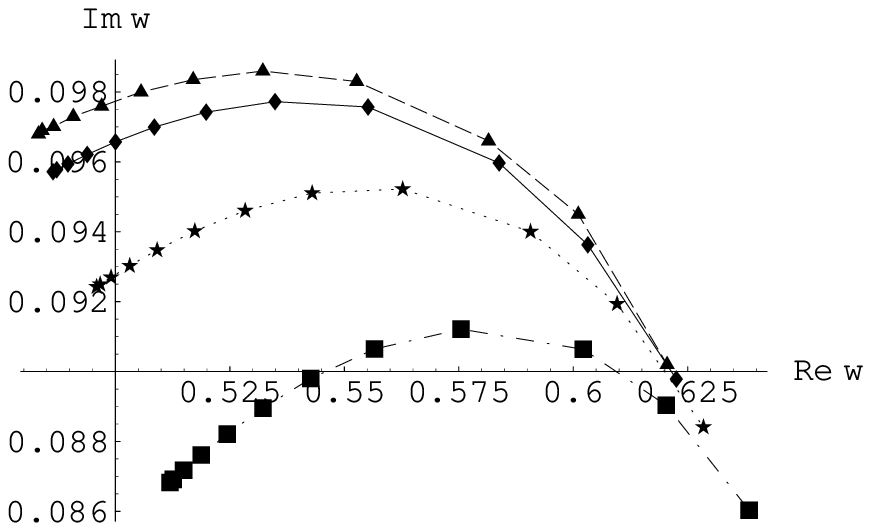}
\caption{RN QNF's for $l=2$, $n=0$, $\mu=0, 0.1, 0.2, 0.3$,
$Q$ runs  the values $0, .1, .2, .3, .4, .5, .6, .7, .8, .9, .95, .99$}
\label{plb1fig2}
\end{center}
\end{figure}

\begin{figure}
\begin{center}
\includegraphics{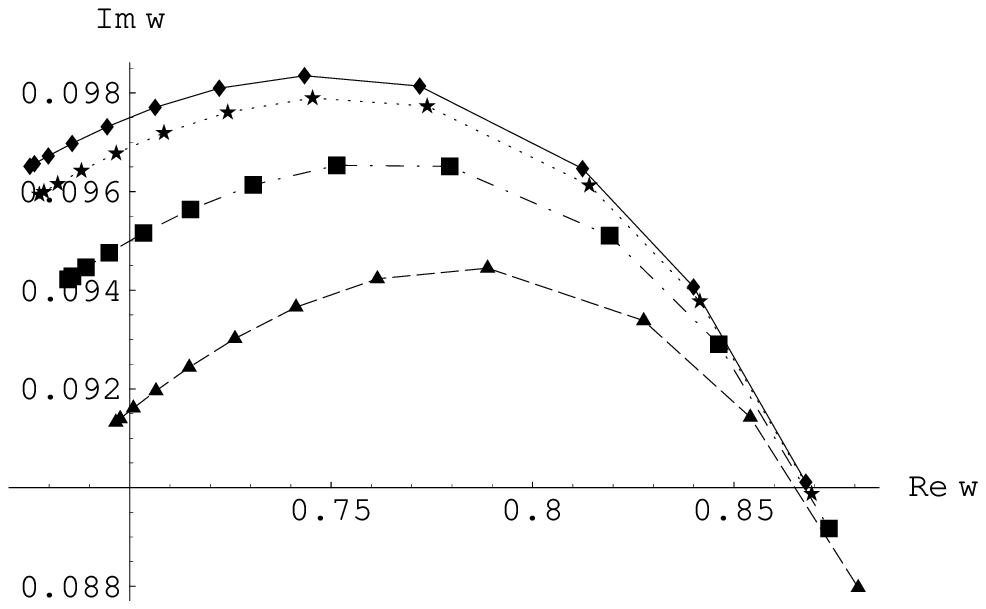}
\caption{RN QNF's for $l=3$, $n=0$, $\mu=0, 0.1, 0.2, 0.3$,
$Q$ runs  the values $0, .1, .2, .3, .4, .5, .6, .7, .8, .9, .95, .99$}
\label{plb1fig3}
\end{center}
\end{figure}

QNF's for different values of $\mu$ are presented on Fig.\ref{plb1fig1}-
\ref{plb1fig3} for $l=1,2,3$. The $l=0$ modes are not given here,
since the WKB accuracy is not sufficient in this case (nevertheless
the main conclusions as to the quasinormal behaviour conserve in this
case as well). The real part of the quasinormal frequency, i.e. the
oscillation
frequency, grows with increasing of the mass field $\mu$, while the
imaginary part of $\omega$, representing the damping rate, falls down.
Note that we are limited to the low-lying quasinormal frequencies
whose real parts correspond to tunneling near the potential barrier
($\omega^2 \approx V_{max}$). In general, the
maximum value of $\mu$ (and of $e$ for charged scalar field)
depends upon the mode under consideration
\cite{Simone-Will}.

Next, we investigate the massive charged scalar field case.
The real part of
$\omega$ as a function of imaginary part of $\omega$ is plotted for
different $l$ and $Q$ for $\mu =0.1$ on Fig.4-6. We see that the neutral
perturbations ($e=0$) have less imaginary part than that of the
charged perturbations. Thus the charged perturbations decay faster.
This repeats the behaviour of the massless case  \cite{Konoplya3}.
When increasing charge $Q$, $\omega_{Im}$ grows up to some maximum
near $Q=0.8 M$ and then falls down, while $\omega_{Re}$ increases
monotonically. The more $e$ the less explicit this maximum.

In addition for the near extremal black hole the imaginary parts of
charged and neutral perturbations tend to coincide as in a massless
case. In other words for the near extremal black hole
$\omega_{Im}$ does not depend on $e$ for a charged massive scalar field.
In this connection if one supposes some kind of relation
between $\omega_{Im}$ of the nearly extremal RN BH and the parameter
of the critical collapse in a massive scalar electrodynamics, then
he could expect that $\gamma_{massive}$ being dependent on mass field
$\mu$ does not depend on $e$ in the region where type II \cite{Gundlach}
critical behaviour occurs.
Yet, at present  we have much more questions then answers here.

\begin{figure}
\begin{center}
\includegraphics{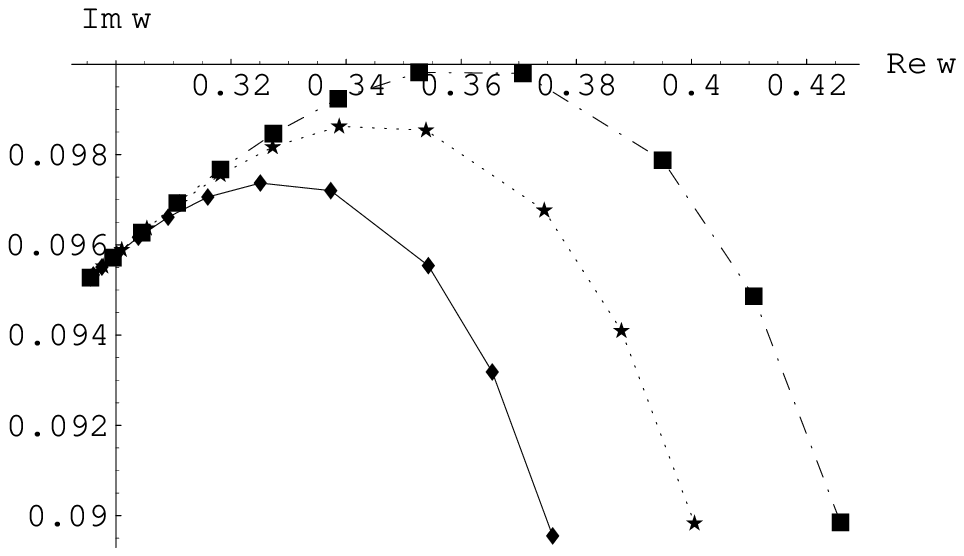}
\caption{RN QNF's for $l=1$, $n=0$, $e=0$ (bottom), $0.05$,
and $e=0.1$ (top),
$Q$ runs  the values $0, .1, .2, .3, .4, .5, .6, .7, .8, .9, .9, .95,
.99$, $\mu =0.1$}
\label{plb1fig4}
\end{center}
\end{figure}

\begin{figure}
\begin{center}
\includegraphics{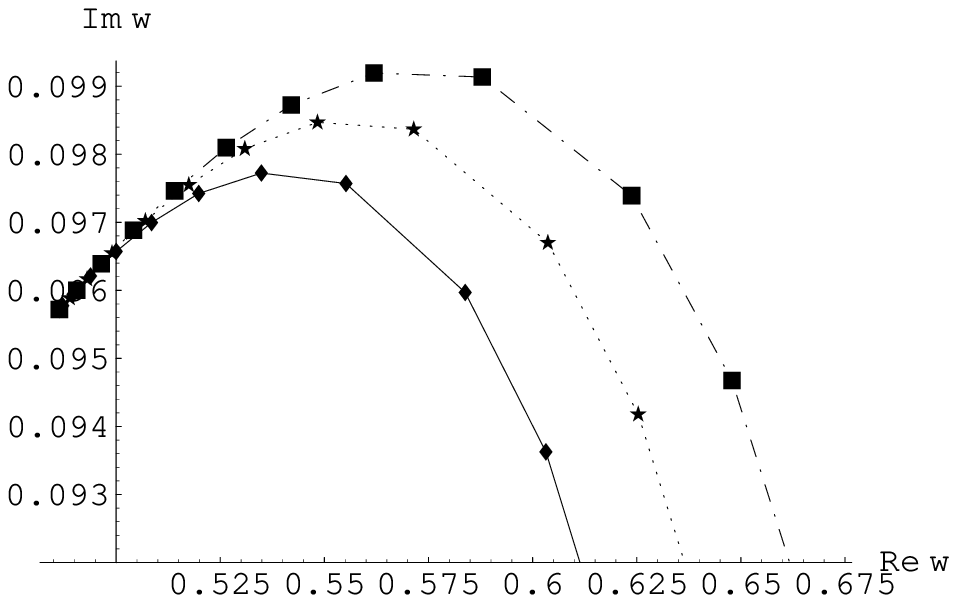}
\caption{RN QNF's for $l=2$, $n=0$, $e=0$ (bottom), $0.05$,
and $e=0.1$ (top),
$Q$ runs  the values $0, .1, .2, .3, .4, .5, .6, .7, .8, .9, .95, .99$,
$\mu =0.1$}
\label{plb1fig5}
\end{center}
\end{figure}

\begin{figure}
\begin{center}
\includegraphics{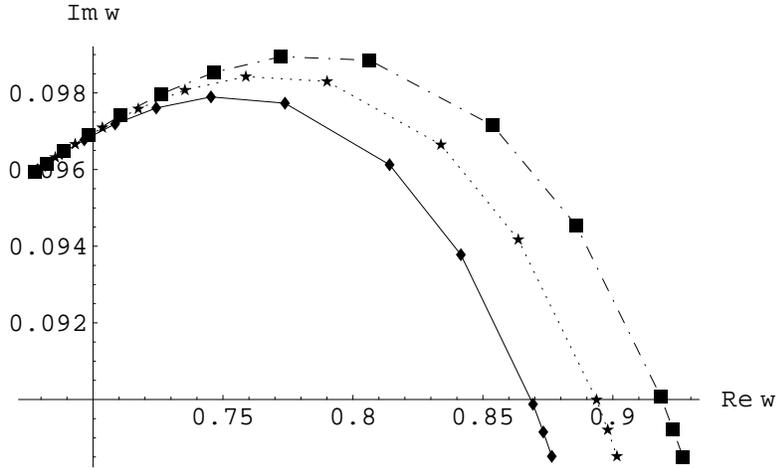}
\caption{RN QNF's for $l=3$, $n=0$, $e=0$ (bottom), $0.05$,
and $e=0.1$ (top),
$Q$ runs  the values $0, .1, .2, .3, .4, .5, .6, .7, .8, .9, .95,
0.995, 0.999$, $\mu =0.1$. At $Q=0.999$ $\omega_{Im}= 0.088517$ at $e=0$,
$\omega_{Im}= 0.088512$ at $e=0.5$, and $\omega_{Im}= 0.088470$ at
$e=0.1$. Thus the coincidence of $\omega_{Im}$ for different $e$ in
the near extremal regime is well within a third order WKB accuracy.}

\label{plb1fig6}
\end{center}
\end{figure}

\end{document}